\documentclass[pdflatex,sn-mathphys]{sn-jnl}

\usepackage{physics}
\usepackage{amsmath,amssymb}
\DeclareMathAlphabet{\pazocal}{OMS}{zplm}{m}{n}

\jyear{2023}%

\theoremstyle{thmstyleone}%
\theoremstyle{thmstyletwo}%

\theoremstyle{thmstylethree}%

\begin{document}

\title[Social Dilemma of Non-Pharmaceutical Interventions]{Social Dilemma of Non-Pharmaceutical Interventions}


\author*[1]{\fnm{Alina} \sur{Glaubitz}}\email{Alina.Glaubitz.GR@dartmouth.edu}

\author[1,2]{\fnm{Feng} \sur{Fu}}\email{Feng.Fu@dartmouth.edu}

\affil*[1]{\orgdiv{Department of Mathematics}, \orgname{Dartmouth College}, \city{Hanover}, \postcode{03755}, \state{NH}, \country{USA}}

\affil[2]{\orgdiv{Department of Biomedical Data Science}, \orgname{Geisel School of Medicine at Dartmouth}, \city{Lebanon}, \postcode{03756}, \state{NH}, \country{USA}}


\abstract{In fighting infectious diseases posing a global health threat, ranging from influenza to Zika, non-pharmaceutical interventions (NPI), such as social distancing and face covering, remain mitigation measures public health can resort to. However, the success of NPI lies in sufficiently high levels of collective compliance, otherwise giving rise to waves of infection incidences that are not only driven by pathogen evolution but also changing vigilance in the population. Here we show that compliance with each NPI measure can be highly dynamic and context-dependent during an ongoing epidemic, where individuals may prefer one to another or even do nothing, leading to intricate temporal switching behavior of NPI adoptions. By characterizing dynamic regimes through the perceived costs of NPI measures and their effectiveness in particular regarding face covering and social distancing, our work offers new insights into overcoming barriers in NPI adoptions.}




\maketitle

\section{Introduction}\label{sec1}

The emergence of new infectious diseases, like Zika~\cite{petersen2016zika}, Ebola~\cite{leroy2005fruit}, COVID-19~\cite{Chams2020} and Mpox~\cite{Banuet2023,Liu2023}, presents a great challenge to global health and humanity. Mathematical models have become an increasingly important part of the understanding and the fight against infectious diseases~\cite{anderson1991infectious}. For an overview see for example Ref. ~\cite{lloyd2017infectious}. These models help inform our response to these threats. Given the time required to develop and distribute pharmaceutical solutions like vaccines, we often have to rely on non-pharmaceutical interventions (NPIs) such as face covering and social distancing in the early stages of an epidemic. However, the effectiveness of these NPIs heavily depends on public compliance and adherence, making human behavior a critical factor in controlling disease spread. Therefore, it is critical to integrate the behavioral response into mathematical models to gain a deeper understanding of the initial spread of infectious diseases.


Notably there has been a growing body of work on the effect of social factors in epidemiology. For an overview of disease-behavior interaction models that have made an impact in this field, see ~\cite{bauch2013social,verelst2016behavioural} and references therein. From the influence of vaccines and vaccine compliance~\cite{Bauch_PLoSCB12} to the role of disease awareness~\cite{Wang_SR16}, these models have made important progress to understanding the role of behavior in epidemiology~\cite{Fu_PRSB11,Reluga_MB06,Wang_PR16}. It has been shown that these interactions can lead to interesting dynamics, like the hysteresis effect~\cite{chen2019imperfect}, by using evolutionary game theory to model how behaviors evolve under the interplay between social influence and self-interest~\cite{bauch2005imitation,Fu_PRSB11,arefin2020vaccinating}. The use of replicator dynamics has been particularly effective in exploring social learning processes and the diffusion of behaviors across various societal challenges, from enforcing norms through peer punishment~\cite{sigmund2001reward} to promoting responsible antibiotic usage~\cite{chen2018social}. In particular, prior work has demonstrated that integrating evolutionary game theory with epidemiological processes can be fruitful in shedding light on disease interventions ~\cite{Fu_PRSB11,chen2019imperfect,Saad-Roy2023,Traulsen2023costs}.

Undoubtedly, the COVID-19 pandemic has led to a further increase of research into the influence of NPIs \cite{Flaxman2020,perra2021non} as well as vaccination strategies \cite{Wagner2021} on epidemics. Unlike pharmaceutical measures like vaccination, NPIs like social distancing and mask-wearing require individuals to continually assess the need to adhere to such measures~\cite{Townsend2020,Saad-Roy2023,Morsky2023}. Thus, it is necessary to account for the socio-economic consequences of the disease and control measures~\cite{nicola2020socio}. Social distancing has been modeled as a game where individuals weigh the risk of the disease against the cost of social distancing to find the Nash equilibrium in the consistent effort to practice social distancing~\cite{reluga2010game}. Optimal control theory has been applied to inform social distancing and lockdown efforts~\cite{maharaj2012controlling,huberts2020optimal}. Additional efforts have been made to understand the effect of social distancing on the COVID-19 epidemic~\cite{aleta2020modelling,kissler2020projecting}, emphasizing the importance of aligning each individual's goals with those of society as a whole.
Further work has focused on modeling vaccination strategies~\cite{Kuga2020,Liu2022}, employing game theory to reveal complex decision-making dynamics. While this previously mentioned work mostly focused on homogeneous populations, additional work has been done in the context of networks, examining the effects of behavioral changes and lockdowns~\cite{Alam2021,Alam2022}, as well as the role of social distancing and adaptive behavior in curbing disease transmission~\cite{Karlson2020, Arthur2021}. Weitz~\cite{Weitz2020} emphasizes the role of awareness in disease mitigation, and Chen~\cite{Chen2022} provides empirical evidence of NPI effectiveness from China's COVID-19 response. Additionally, work has been done to understand how face covering and asymptomatic transmission changes the spread of infectious diseases ~\cite{Qiu2022,Espinoza2021,Espinoza2022}.

Moreover, optimal control theory has been applied to model the implementation of mandated vaccination and lockdowns, aiming to minimize outbreak costs through centralized planning~\cite{sethi1978optimal,abakuks1974optimal,abakuks1973optimal,wickwire1975optimal}. Despite providing valuable insights into achieving population-wide optimal outcomes~\cite{Morris2021}, the practical implementation of these strategies can be limited by failing to ensure widespread compliance. Our previous work ~\cite{glaubitz2020oscillatory} emphasizes the role of individual compliance and identifies an ``oscillatory tragedy of the commons" in social distancing dynamics~\cite{glaubitz2020oscillatory} by implementing bounded rationality~\cite{simon1990bounded} and loss aversion~\cite{tversky1992advances}.

This work builds upon our prior study ~\cite{glaubitz2020oscillatory} and explores the intricate effects of a dual behavioral response on pandemic responses. Expanding the previous model, we introduce face covering (FC) as a less strict form of NPI compared to social distancing. FC serves as a second type of intervention strategy, contributing to a `Swiss cheese' model of protection. Our findings reveal that FC as a third strategy next to social distancing (SD) and no social distancing (NSD) displays a significantly different behavior than SD. While SD is generally preferred over NSD once a threshold is passed, FC is preferred over NSD only within a range between two disease prevalence thresholds. Once the infections pass the smaller threshold, FC is preferred, but once the second threshold is passed, NSD is preferred over FC again. Additionally, FC is only adopted if its effectiveness is large enough compared to SD. Once the relative effectiveness of FC drops below a threshold, it does not get adopted at all. Moreover, we show how individuals generally opt for either SD or FC, without reverting to more stringent measures (SD) once FC is preferred.

\section*{Model and Methods}

We incorporate into an epidemiological model the human behavior choices of NPI measures, using the replicator dynamics to account for the decision whether to SD, FC, or NSD. This significantly extends our prior work \cite{glaubitz2020oscillatory}  and results in a `Swiss cheese' model of protection with a dual behavioral response. 

Each individual is either susceptible (S), infected (I) or recovered (R). Those who are susceptible have a choice in how they respond to the threat of infection: they can practice SD, FC, or NSD. A susceptible without face covering gets infected in an encounter with an infected with rate $\beta$, a susceptible with face covering gets only infected with rate $\beta_{\text{FC}} < \beta$. At any given moment, an infected recovers with rate $\gamma$. To account for the force of infection, we use a well-mixed population model. Extending this model to networked populations is straightforward~\cite{Qiu2022}. 

\begin{figure}[t]
    \centering
    \includegraphics[width=.8\linewidth]{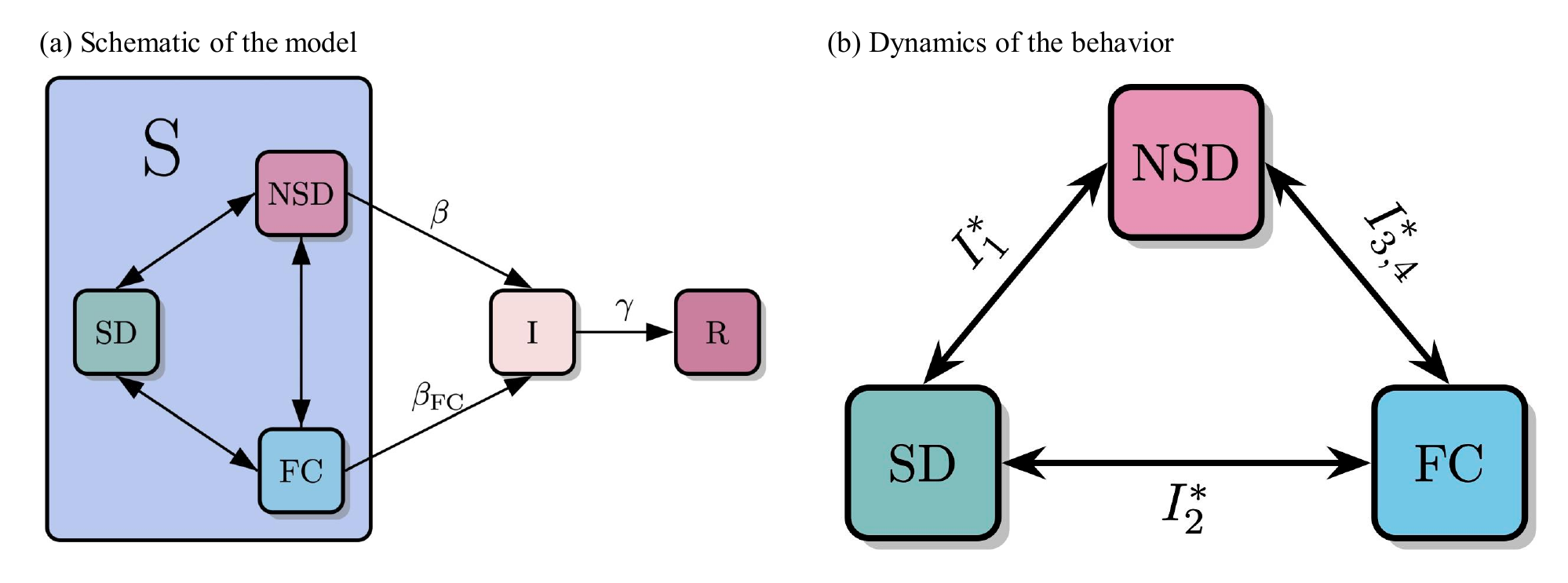}
\caption{Model schematic. 
(a) Overview of the model: The susceptible population is divided into three groups here: those practicing social distancing (SD), those practicing no social distancing but wearing a face covering (FC), and those neither practicing social distancing nor wearing face covering (NSD). Everyone not practicing social distancing can be infected, though face covering reduces the risk from $\beta$ to $\beta_{\text{FC}}$. The dynamics of the strategies within the susceptible population are governed by replicator equations. (b) Disease-prevalence dependent behavior adoption dynamics. Individuals' adoption preference of NPIs changes as a result of the prevalence of infection. In particular, which NPI measure is preferred or inaction at all depends on whether $I$ is smaller or larger than the three prevalence thresholds. Note that multiple thresholds can arise between NSD and FC, indicating possible multistability and bifurcation.}
\label{fig:schematic}
\end{figure}

We denote the proportion of susceptible individuals at time $t$ by $S(t)$, the proportion of infected by $I(t)$, and the proportion of removed by $R(t)$. Furthermore, we denote by $\mathcal{E}(t)$ the proportion of susceptible individuals that practice SD, by $\mathcal{F}(t)$ the proportion of people in FC and by $\mathcal{G}(t)$ the proportion of in NSD. The initial conditions we denote by $I_0 = I(0), S_0=S(0)$ as well as $\mathcal{E}_0 = \mathcal{E}(0), \mathcal{F}_0 = \mathcal{F}(0), \mathcal{G}_0 = \mathcal{G}(0)$.

A susceptible individual updates their NPI choices (behavioral strategies consisting of NSD, SD, and FC) based on a simple introspection of the costs associated with each NPI choice. Thus,
by $\pi_{\text{SD}}$ we denote the payoff of social distancing, by $\pi_{\text{FC}}$ the payoff of wearing a face covering and by $\pi_{\text{NSD}}$ the payoff of no social distancing and no face covering. Social distancing in our model has cost $C_{\text{SD}}>0$ at each time $t$. Thus, we have 
\[
    \pi_{\text{SD}} = - C_{\text{SD}}.
\]
$\pi_{\text{NSD}}$ depends on two factors: the cost of infection that we denote by $C_{\text{I}}>0$ and the risk of infection. The risk of infection in time $(t,t+1)$ when NSD is given by 
\[
    1 - \exp\left(-\beta \int_t^{t+1} I(\tau) \mathrm{d}\,\tau\right) \approx 1 - \exp\left(-\beta I(t)\right).
\]
Thus, the payoff of NSD is given by
\[
    \pi_{\text{NSD}} = -C_{\text{I}} (1-\exp\left(-\beta I(t)\right)).
\]
Wearing face covering has cost $C_{\text{FC}}$ and analogous to $\pi_{\text{NSD}}$ we therefore have
\[  
    \pi_{\text{FC}} = -C_{\text{FC}} - C_{\text{I}}\left( 1-\exp\left( -\beta_{\text{FC}} I(t) \right) \right)
\]
The dynamics of our model are given by the following system of ODEs:
\begin{align}\label{eq:SIR_FC}
\begin{split}
        \dot{S}(t) &= -\beta \mathcal{G}(t) S(t) I(t) - \beta_{\text{FC}} \mathcal{F}(t) S(t) I(t) \\
        \dot{I}(t) &= \beta  \mathcal{G}(t) S(t) I(t) + \beta_{\text{FC}} \mathcal{F}(t) S(t) I(t) - \gamma I(t) \\
        \dot{R}(t) &= \gamma I(t)
\end{split}
\end{align}
as well as 
\begin{align} \label{eq:SD_FC_dyn}
\begin{split}
    \dot{\mathcal{E}}(t) &= \omega \mathcal{E}(t)\mathcal{F}(t)\tanh\left(\frac{\kappa}{2}\left(-C_{\text{SD}} + C_{\text{FC}} + C_I (1 - e^{-\beta_{\text{FC}} I(t)})\right)\right) \\ &\qquad+ \omega \mathcal{E}(t)\mathcal{G}(t)\tanh\left(\frac{\kappa}{2}(-C_{\text{SD}} + C_I (1 - e^{-\beta I(t)}))\right) \\
        \dot{\mathcal{F}}(t) &= \omega\mathcal{F}(t)\mathcal{E}(t)\tanh\left(\frac{\kappa}{2}(-C_{\text{FC}} - C_I (1 - e^{-\beta_{\text{FC}} I(t)})\right) \\ &\qquad+ \omega \mathcal{F}(t)\mathcal{G}(t)\tanh\left( \frac{\kappa}{2}(-C_{\text{FC}}+  C_I ( e^{-\beta_{\text{FC}} I(t)} - e^{-\beta I(t)}))\right) \\
        \dot{\mathcal{G}}(t) &= \omega \mathcal{G}(t)\mathcal{E}(t)\tanh\left(\frac{\kappa}{2}(-C_I (1 - e^{-\beta_{\text{FC}} I(t)}) + C_{\text{SD}})\right) \\ &\qquad+ \omega \mathcal{G}(t)\mathcal{F}(t)\tanh\left(\frac{\kappa}{2}(C_{\text{FC}} + C_I (e^{-\beta I(t)} - e^{-\beta_{\text{FC}} I(t)}))\right)
\end{split}
\end{align}
Here, $\omega$ is a responsiveness parameter and $\kappa$ is a rationality parameter. Together, these parameters determine how fast the susceptible population adapts their NPI strategies to change in the prevalence of infected $I$ (see the Supplementary Information for details). 

\section*{Results}\label{sec:results}

The adoption behavior dynamics of the model are governed by three prevalence thresholds $I_n^*$:
\begin{align}
    \pi_{\text{SD}} = \pi_{\text{NSD}} &\iff I_1^* = - \frac{1}{\beta} \log \left( 1 - \frac{C_{\text{SD}}}{C_{\text{I}}} \right) \nonumber\\
    \pi_{\text{SD}} = \pi_{\text{FC}} &\iff I_2^* = - \frac{1}{\beta_{\text{FC}}} \log \left( 1 - \frac{C_{\text{SD}}-C_{\text{FC}}}{C_{\text{I}}} \right) \nonumber \\
    \pi_{\text{FC}} = \pi_{\text{NSD}} &\iff \frac{C_{\text{FC}}}{C_{\text{I}}} = \exp(-\beta_{\text{FC}} I_3^*) - \exp(-\beta I_3^*). \label{eq:i3}
\end{align}
Simply put, when the infection rate ($I$) is below these thresholds ($I<I_n^*, n \in {1,2}$), people find it more beneficial not to engage in social distancing or face covering. They lean towards 'normal behavior' (NSD) or using face coverings as a lighter precaution (FC) compared to SD. However, once the infection rate crosses these thresholds ($I>I_n^*, n \in {1,2}$), SD becomes the more attractive option due to its higher benefits in preventing disease spread.

\begin{figure}[t]
    \centering
    \includegraphics[width=.8\linewidth]{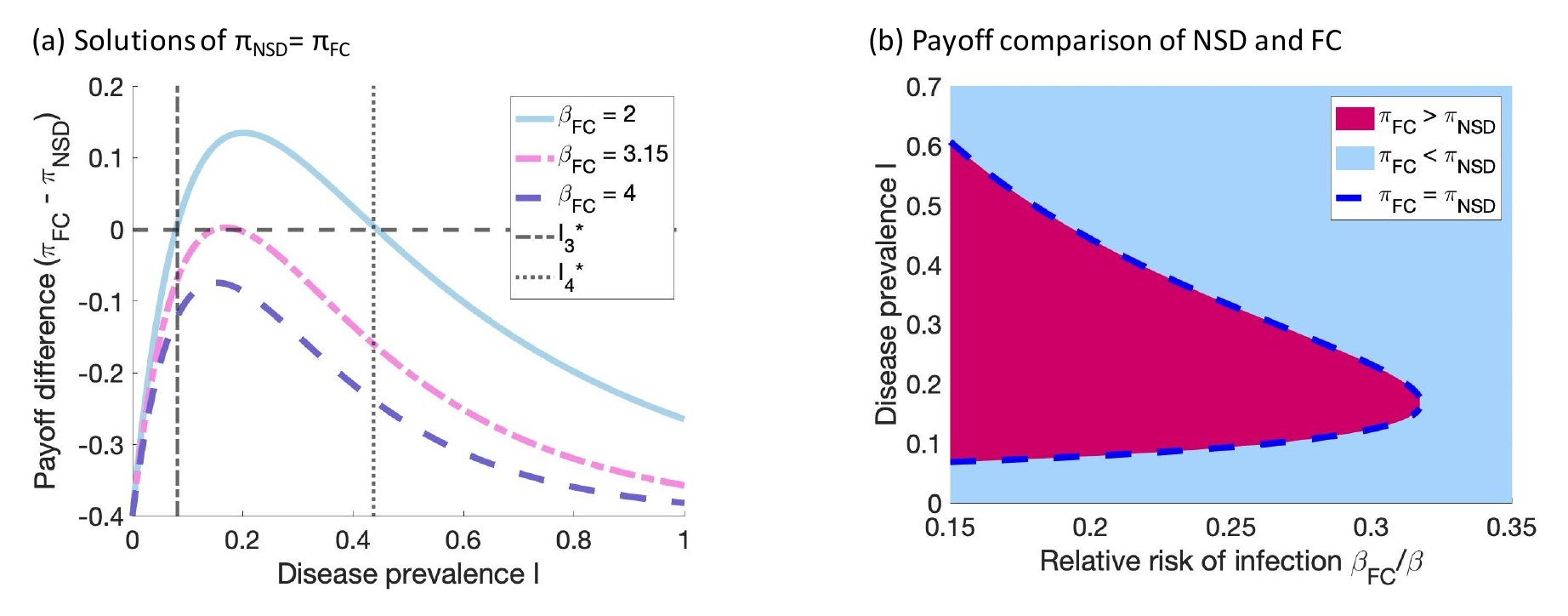}
\caption{Efficacy of face covering impacts its adoption. (a) Shows examples of different efficacies of face covering for $\pi_{\text{FC}}-\pi_{\text{NSD}}$ as a function of $I$. Model parameters:$\beta = 10$, $C_{\text{I}} = 5$ and $C_{\text{FC}} = 2$. (b) Shows the parameter regions $(\beta_{\text{FC}},I)$ where $\pi_{\text{FC}} > \pi_{\text{NSD}}$ (red), $\pi_{\text{FC}} = \pi_{\text{NSD}}$ (dashed) and $\pi_{\text{FC}}<\pi_{\text{NSD}}$ (blue).}
\label{fig:I3}
\end{figure}

The situation gets more complex with the third threshold, which involves a comparison between not practicing any social distancing (NSD) and opting for lighter measures (FC). This threshold, defined by two possible points, $I_3^*$ and $I_4^*$, creates a scenario where for disease prevalence levels $I(t)$ between these two points, wearing face coverings is more advantageous than not doing anything at all. Specifically, for $I(t)$ lower than $I_3^*$, people don’t see the need for face coverings. Between $I_3^*$ and $I_4^*$, face coverings are adopted as they offer a better payoff compared to doing nothing. However, once $I(t)$ surpasses $I_4^*$, the incentive to continue wearing face coverings diminishes, and people stop using them. Furthermore, if FC is not effective enough compared to SD, ($\beta_{\text{FC}}$ too large compared to $\beta$), neither threshold $I_n^*, n \in \{3,4\},$ exists and FC is never preferred over NSD. (See Figure \ref{fig:I3} for an illustration.)

To determine when FC cannot be the favored strategy anymore, we define $F(x) = \exp(-\beta_{\text{FC}} x) - \exp(-\beta x)$ as the RHS of \eqref{eq:i3}. This function attains its maximum for $$x_{\text{max}} = \frac{\log(\beta)-\log(\beta_{\text{FC}})}{\beta - \beta_{\text{FC}}}$$.

When the peak of this function is higher than the LHS of \eqref{eq:i3} ($\frac{C_{\text{FC}}}{C_{\text{I}}}$), it means that there are two specific prevalence levels, $I_3^*$ and $I_4^*$, between which FC is preferred over NSD. If the peak is lower than this ratio, it suggests that face coverings do not offer a better payoff compared to NSD, essentially making them undesired.

To find out the least effectiveness level of face coverings (expressed as $\beta_{\text{FC}}$) for them to be considered a better strategy, we calculate the maximum value of $F(x)$ and equate it to $\frac{C_{\text{FC}}}{C_{\text{I}}}$. Solving this equation gives us a critical threshold $\beta_{\text{FC}}^*$ as the solution of
\begin{equation}
    \label{eq:beta_fc*}
    \exp\left(-\beta_{\text{FC}} \frac{\log(\beta)-\log(\beta_{\text{FC}})}{\beta - \beta_{\text{FC}}}\right) - \exp\left(-\beta \frac{\log(\beta)-\log(\beta_{\text{FC}})}{\beta - \beta_{\text{FC}}}\right) = \frac{C_{\text{FC}}}{C_{\text{I}}},
\end{equation} 
indicating the minimum effectiveness level required for face coverings to be beneficial. If the effectiveness of face coverings falls below this threshold, they are not considered a preferred strategy. This analysis helps us pinpoint the conditions under which individuals decide between wearing face coverings and not doing anything to prevent infection spread, providing a clearer picture of understanding NPI behavior choices during epidemics.

\begin{figure}[t]
    \centering
    \includegraphics[width=.8\linewidth]{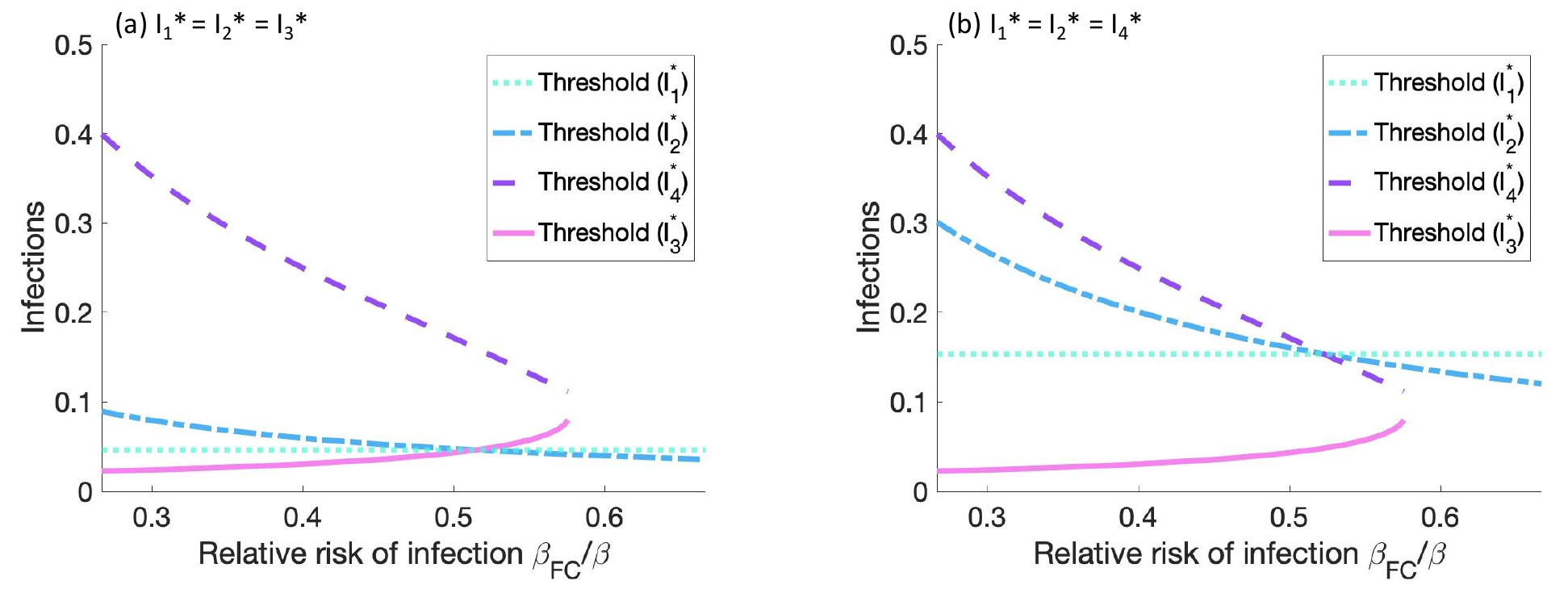}
\caption{Order of disease prevalence thresholds. Effective face coverings always enhance its adoption compared to social distancing. But low efficacy of face covering will likely lead to low adoption unless social distancing is sufficiently costly. Model parameters: $\beta = 15$, $C_{\text{FC}}=10$, $C_{\text{I}}=50$ and different $C_{\text{SD}} = 25; 45$.}
\label{fig:beta_fc_thresholds}
\end{figure}

Additionally, if $I_1^*=I_2^*$, this implies that either $I_1^* = I_2^* = I_3^*$ or $I_1^* = I_2^* = I_4^*$. To see why that is, note that
$$
    \exp(-\beta I_1^*) = 1 - \frac{C_{\text{SD}}}{C_{\text{I}}}, \text{ and } \exp(-\beta_{\text{FC}} I_2^*) = 1 - \frac{C_{\text{SD}}-C_{\text{FC}}}{C_{\text{I}}}.
$$
For $I = I_1^* = I_2^*$, this implies 
$$
    F(I) = \exp(-\beta_{\text{FC}} I_1^*) - \exp(-\beta I_2^*) = \frac{C_{\text{FC}}}{C_{\text{I}}}.
$$
Therefore, $I_1^*$ then solves \eqref{eq:i3} and if $I_1^* = I_2^*$ holds then we either have $I_1^* = I_2^* = I_3^*$ or $I_1^* = I_2^* = I_4^*$. An illustration of these cases can be seen in Figure \ref{fig:beta_fc_thresholds}. Note that we will ignore the null set where $I_1^* = I_2^*$ or $I_3^* = I_4^*$ for this analysis.

In the first scenario, where we compare whether $I_1^* = I_2^* = I_3^*$ holds or does not hold, the thresholds can only be ordered in one of the following ways:
\begin{align*}
    \text{(i) } I_2^* < I_1^* < I_3^*<I_4^*, \quad
    \text{(ii) } I_3^* < I_1^* < I_2^*<I_4^*, \\
    \text{(iii) } I_2^*<I_1^* \text{ and neither } I_3^* \text{ nor } I_4^*  \text{ exists.}
\end{align*}

In the second scenario, where we compare whether $I_1^* = I_2^* = I_4^*$ holds or does not hold, we can have 
\begin{align*}
    \text{(ii) } I_3^* < I_1^* < I_2^*<I_4^*, \quad
    \text{(iii) } I_2^*<I_1^* \text{ and neither } I_3^* \text{ nor } I_4^*  \text{ exists,} \\
    \text{(iv) } I_3^*<I_4^* < I_2^*<I_1^*.
\end{align*}

\begin{figure}[t]
    \centering
    \includegraphics[width=.8\linewidth]{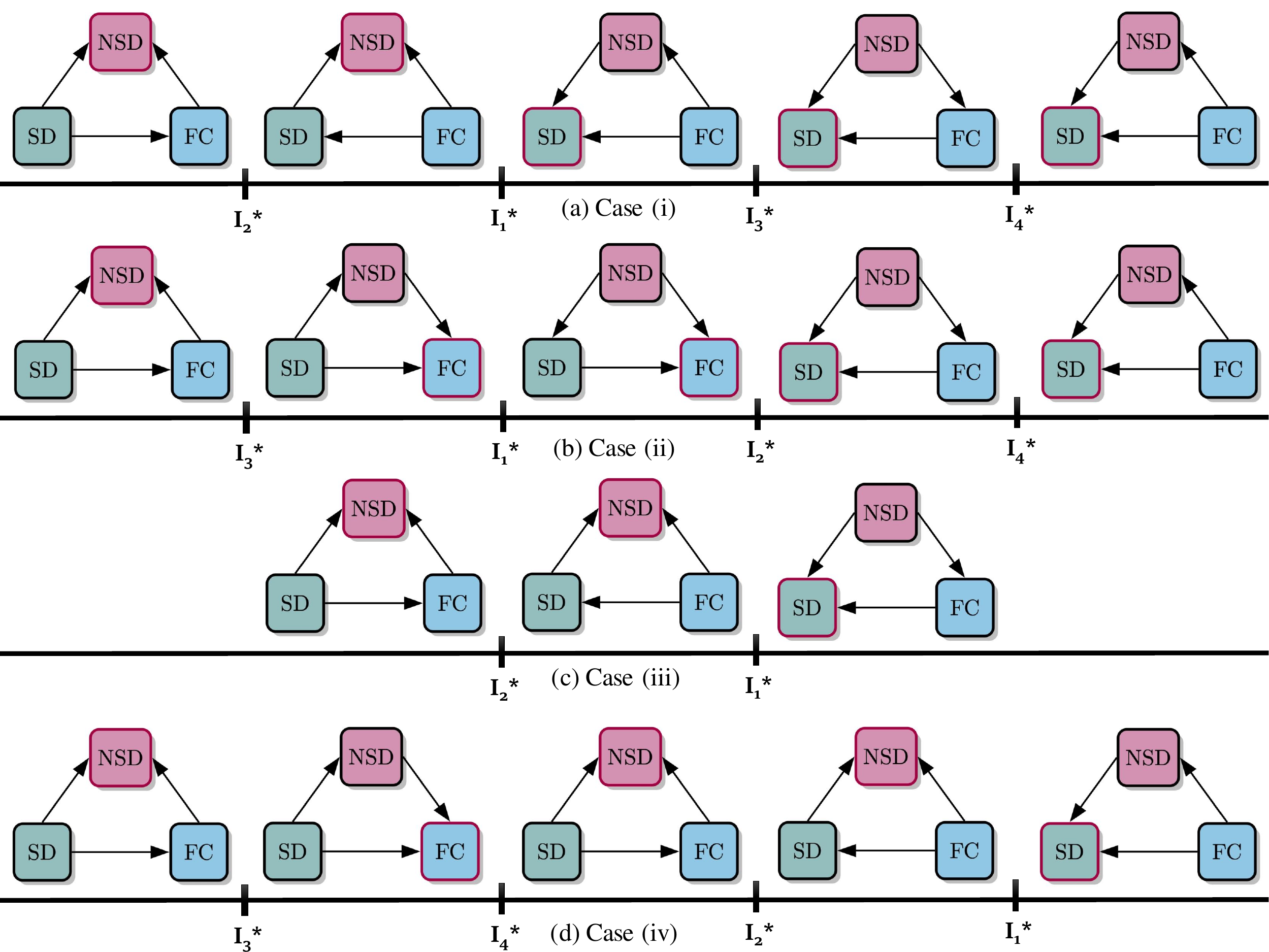}
\caption{Context-dependent adoption preference of NPI measures. The dynamics of the different strategies. For very small prevalence of infected $I$, the dynamics always tend towards taking no measures. For very large prevalence of infected $I$, the dynamics tend towards social distancing. How they behave in between depends on the order of the four thresholds. Predominant strategies are highlighted with bold red borders.}
\label{fig:dynamics}
\end{figure}

To understand the adoption dynamics of NPI measures in each of the cases, see Figure \ref{fig:dynamics}. Here we see the order of predominant strategies for all four cases. In case (i), the infection is mitigated through SD similar to the SIR-SD model \cite{glaubitz2020oscillatory}. In case (ii), FC is the predominant strategy for mitigation, but players might also practice SD if the infection becomes more prevalent. In particular, we might see SD in the first stages/oscillations of the epidemic here and FC in the later stages. In case (iii), FC is never adopted by the population and SD is adopted in the first scenario, but not the second. In case (iv), only FC will ever be adopted as players stop FC before SD ever gets adopted.

\begin{figure}[t]
    \centering
    \includegraphics[width=.8\linewidth]{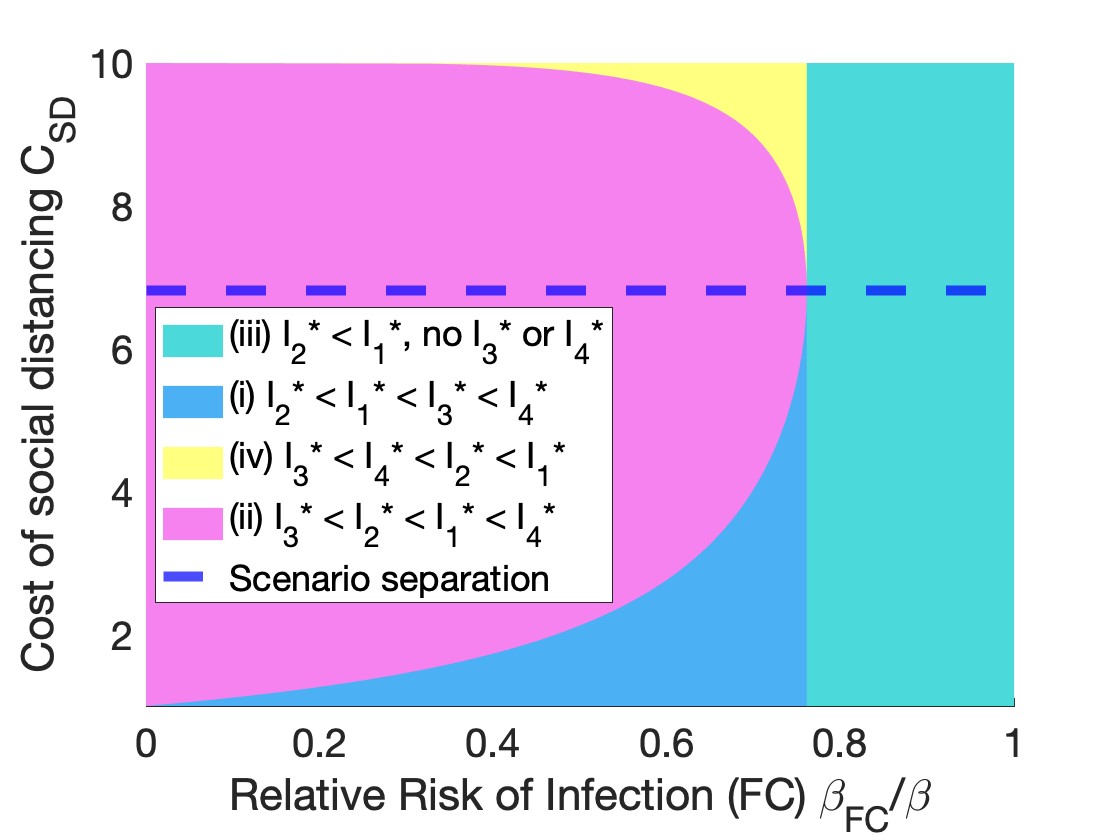}
\caption{2D phase plot of the order of the four thresholds with $\beta = 1, C_{\text{I}}=10$ and $C_{\text{FC}}=1$ and illustration for which scenario occurs. Here, we determine the specific combinations of $\beta_{\text{FC}}$ and $C_{\text{SD}}$ for each of the resulting scenarios, respectively, along with the order of the prevalence thresholds.}
\label{fig:phase_plot}
\end{figure}

Next, we want to derive the condition under which these scenarios occur. 
For the four thresholds to align, we need
$
    \beta_{\text{FC}}^*/\beta = \log(1- \frac{C_{\text{SD}}-C_{\text{FC}}}{C_{\text{I}}})/\log(1- \frac{C_{\text{SD}}}{C_{\text{I}}})
$
to be satisfied. Then, when fixing $C_{\text{I}}, \beta, C_{\text{FC}}$, we have a threshold $C_{\text{SD}}^*$ such that for $C_{\text{SD}} < C_{\text{SD}}^*$, the first scenario occurs and for $C_{\text{SD}} > C_{\text{SD}}^*$ the second scenario occurs. 
An illustration of this is given in Figure \ref{fig:phase_plot}.

\begin{figure}[t]
    \centering
    \includegraphics[width=.8\linewidth]{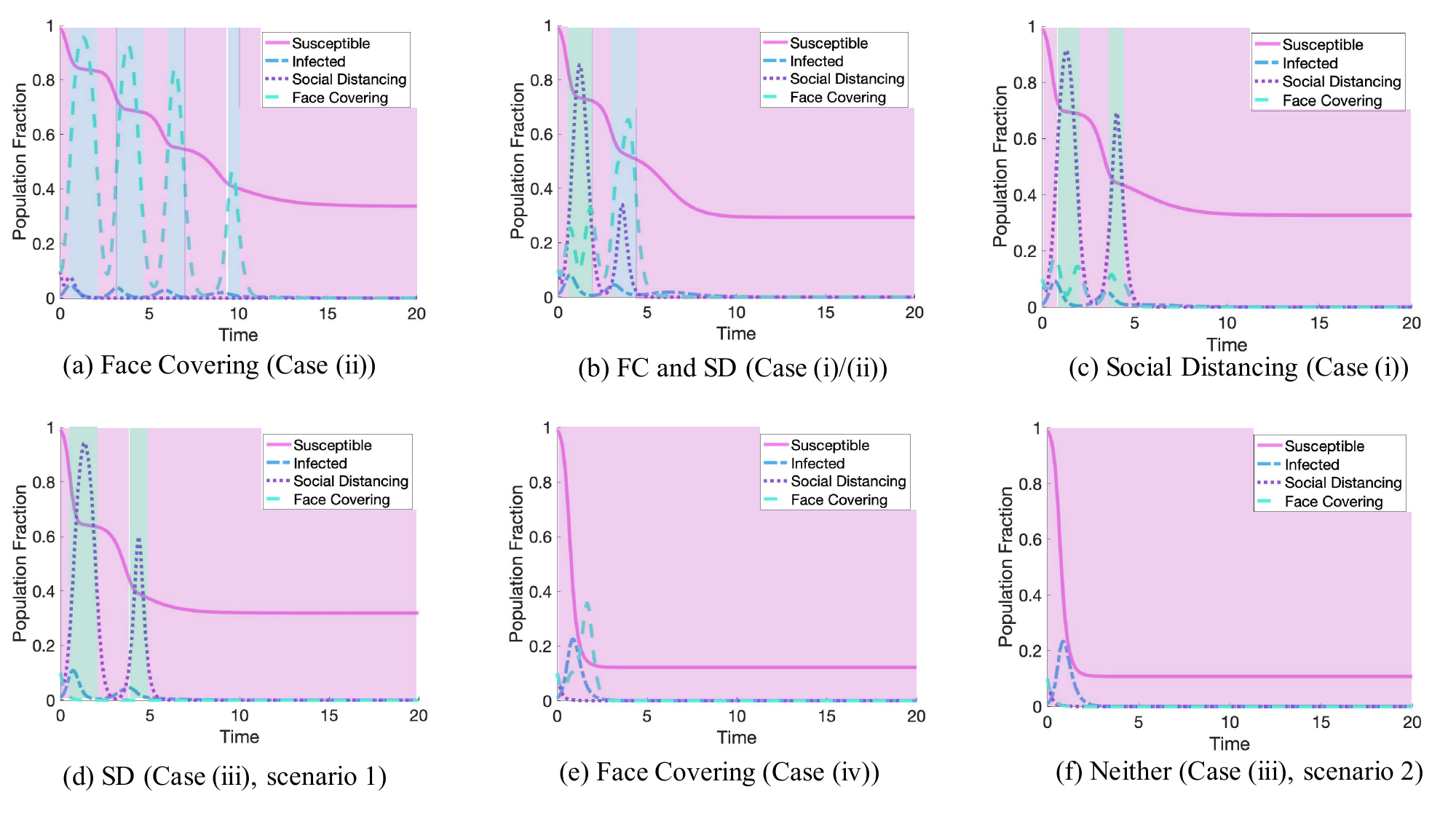}
\caption{Dynamic adoption preferences and compliance to each NPI, including social distancing and face covering. For very small disease prevalence, the dynamics tend towards no measures (NSD), and for very large disease prevalence towards social distancing (SD). In between, they tend towards face covering for some cases. For each case, the time window during which these exist a preferred NPI measure (FC vs SD) is highlighted by the corresponding shaded color. Model parameters: $\beta = 10$, $\gamma = 4$, $\omega=5$, $\kappa = 5$, $C_{\text{I}}=10$, $C_{\text{FC}}=1$ and (a) $C_{\text{SD}}= 2$, $\beta_{\text{FC}} = 1$,  (b) $C_{\text{SD}}= 2$, $\beta_{\text{FC}} = 4$, (c) $C_{\text{SD}}= 2$, $\beta_{\text{FC}} = 5$, (d) $C_{\text{SD}}= 2$, $\beta_{\text{FC}} = 8.5$, (e) $C_{\text{SD}}= 8.5$, $\beta_{\text{FC}} = 7$ and (f) $C_{\text{SD}}= 8.5$, $\beta_{\text{FC}} = 8.5$.}
\label{fig:example}
\end{figure}

So, there are six different cases to be discussed, as demonstrated in Figure~\ref{fig:example}:

\begin{itemize}
    \item[Case \textbf{(ii)}] \textbf{High Efficiency of FC:} If FC is efficient ($\beta_{\text{FC}}/\beta$ small enough), face covering will be the preferred strategy and dampen the infection dynamics (magenta area in Figure \ref{fig:beta_fc_thresholds}, illustration in Figure \ref{fig:example} (a)).
    \item[Case \textbf{(ii)/(i)}] \textbf{Transition from SD to FC:} As FC becomes less efficient (boundary of the magenta and blue area) and the cost of SD is still small enough, SD is followed by FC as a measure to slow down the infection (Figure \ref{fig:example} (b)).
    \item[Case \textbf{(i)}] \textbf{SD Dominance with FC Remnants:} For increasingly less efficient FC (blue area), there are still some remainders of FC, but SD is now predominantly controlling the disease dynamics (Figure \ref{fig:example} (c)).
    \item[Case \textbf{(iii)}] \textbf{Exclusive Use of SD:} In scenario 1, once FC is not efficient enough to provide any advantage (teal area), only SD as a measure is taken (Figure \ref{fig:example} (d)).
    \item[Case \textbf{(iv)}] \textbf{Limited Use of FC:} If FC is less efficient, but at the same time SD is very costly (yellow area), only a small level of FC is practiced (Figure \ref{fig:example} (e)).
    \item[Case \textbf{(iii)}] \textbf{No Measures Adopted:} In scenario 2, once FC does not provide any advantage anymore (teal area above scenario separation), players do not take any measures against infection (Figure \ref{fig:example} (f)).
\end{itemize}

These scenarios highlight a critical insight: once the population transitions to using face coverings due to their efficiency or as a compromise between no measures and the high cost of SD, there is no return to more stringent measures. This shift indicates a one-way trajectory in public health behavior during an epidemic, where once less stringent measures like FC are adopted due to their perceived efficiency or cost-effectiveness, the population does not revert to more restrictive interventions like SD, even as the dynamics of the infection and the relative efficiency of interventions evolve.

\section*{Discussion}

This work extends upon previous studies on NPIs in infectious disease control by incorporating a dual behavioral response (SD and FC) compared to models typically with only one possible behavioral response for mitigating the spread of an infectious disease. However, the model also presents certain limitations that should be discussed.

Firstly, the model does not account for factors such as seasonality~\cite{Wiemken2023}, the emergence of new strains through mutation~\cite{Tanaka2002}, age-related differences in infection risk~\cite{Arthur2023}, or heterogeneity in the costs associated with social distancing or face covering~\cite{Espinoza2021}. Moreover, the influence of asymptomatic individuals~\cite{Espinoza2022} and vaccination~\cite{Jentsch2021} on the dynamics of disease spread and control measures are not considered. These factors could significantly impact the effectiveness and societal acceptance of NPIs.

Another limitation lies in the assumption that the costs of social distancing and mask-wearing are constant over time. In reality, as the pandemic evolves, public perception and tolerance of these measures may shift, leading to changes in compliance levels~\cite{Shearston2021}. Additionally, the model does not incorporate top-down governmental interventions such as mask mandates or lockdown orders, which can play a crucial role in shaping population behavior.

The complexity of human behavior in response to the infection risks and intervention costs associated with infectious diseases has been investigated in related works. It has been shown that age differences in the cost of infection can introduce chaotic behavior into epidemiological models~\cite{Arthur2023}. Previous studies have approached mask-wearing and social distancing from network \cite{Arthur2023} and game-theoretical perspectives \cite{Traulsen2023costs}. Prior work has also incorporated  vaccination \cite{Saad-Roy2023} and threshold-dependent tipping dynamics \cite{Morsky2023} into models of social distancing behavior, highlighting the complex interplay between various measures and the spread of infection.

Our work builds upon these contributions by introducing a model that considers two sorts of NPI measures — SD and FC — and their interaction. This approach allows for a more detailed exploration of the strategic decisions individuals make in the face of an epidemic and provides insights into how multiple concurrent measures (the so-called `Swiss cheese' model of protection) can influence the trajectory of disease spread. The model reveals that once less stringent measures like FC are adopted due to their perceived benefits, populations do not revert to more restrictive interventions like SD, suggesting a one-way transition trajectory in NPI adoption behavior.

In conclusion, our model offers new insights into the dynamics of NPI adoption and compliance, particularly emphasizing that their context-dependent adoption preferences should be taken into account by top-down public health interventions. By extending the current work, future research aiming to overcome barriers in NPI adoptions can incorporate a broader range of factors that influence human behavior and disease spread. Understanding the interaction between different sorts of preventive measures and their impact on epidemiological outcomes will be crucial for designing effective public health strategies in response to current and future pandemics.

\backmatter

\bmhead{Supplementary information}

If your article has accompanying supplementary file/s please state so here. 

Authors reporting data from electrophoretic gels and blots should supply the full unprocessed scans for key as part of their Supplementary information. This may be requested by the editorial team/s if it is missing.

Please refer to Journal-level guidance for any specific requirements.

\bmhead{Acknowledgements}

We are grateful for the support by the Bill \& Melinda Gates Foundation (award no.~OPP1217336), the NIH COBRE Program (grant no.~1P20GM130454), and the Burke Research Award.

\noindent
If any of the sections are not relevant to your manuscript, please include the heading and write `Not applicable' for that section. 


\begin{appendices}

\section{Perfect Adoption}

An important factor in the SIR-SD-FC model is the assumption of bounded rationality. The parameters $\omega$ and $\kappa$ in the model ensure that the model displays waves of infection. For $\kappa,\omega \to \infty$, the population can react to each shift in the optimal strategy infinitely fast. This leads to the population always oscillating in smaller and smaller waves around the smaller of the thresholds $I_k^*, k \in \{1,3\}$ until the population reaches herd immunity. So, there are two cases: (a) For relatively effective face covering (denoted as case (ii)/case(iii) in the main text), the fully rational population uses only face covering to fix the population at the threshold $I_3^*$. (b) For less effective face covering (case (i)/case(iv)), the proportion of infected is fixed around $I_1^*$ until we reach herd immunity. For both cases, we obtain similar dynamics, given by the following ODEs
\begin{align}\label{eq:perfect_adaption}
\begin{split}
    \dot{I}_{\text{PA}}(t) &= \begin{cases}
            0,&  t < t^* \\
            \beta I(t) S(t) - \gamma I(t), & t>t^*
        \end{cases} \\
    \dot{R}_{\text{PA}}(t) &= \gamma I(t)
\end{split}
\end{align}
with $S_{\text{PA}}(t) = 1 - I_{\text{PA}}(t) - R_{\text{PA}}(t)$, $t^* = - \frac{\gamma - \beta + \beta I^*_k}{\beta I^*_k \gamma}, k \in \{1,3\}$, and with initial condition 
\[
    I_{\text{PA}}(0) = I^*_k, \qquad R_{\text{PA}}(0) = 0.
\]
Then, the total amount of people that get infected $R_{\text{PA}}(\infty)$ is given by
\[
    R_{\text{PA}}(\infty) = \frac{\gamma}{\beta} W\left( - \exp\left( -\frac{I^*_k \beta}{\gamma} - 1 \right) \right) + 1,
\]
where $W$ denotes the Lambert $W$ function.
Thus, in the case of perfect adoption, we can achieve
\[
    R_{\text{PA}}(\infty) \to 1 - \frac{\gamma}{\beta}
\]
for the relative cost of social distancing and face covering compared to infection $\frac{C_{SD}}{C_I} \to 0$ or $\frac{C_{FC}}{C_I} \to 0$. In particular, the model displays a similar behavior to the SIR-SD model in this scenario. We only either have SD or FC happening to mitigate the disease spread whenever we have a perfectly rational population as we do not observe overshooting of the infections over the lower threshold.

For perfect adoption, when varying the cost of SD, the total number of infections monotonically increases until the threshold for face covering is smaller than the threshold for social distancing (i.e. $I_3^*<I_1^*$). At this point, face covering becomes the strategy of choice and the total fraction of infected $R(\infty)$ becomes constant. For expensive and/or ineffective face covering, this means that $R(\infty)$ increases until we reach the SIR model baseline of infections. For the SIR-SD-FC model (denoted as bounded rationality in the figure), we observe oscillations similar to the SIR-SD model until the total number levels off less than or equal to the perfect adoption model. For an explanation, see \cite{glaubitz2020oscillatory}. If, on the other hand, we vary the effectiveness of face covering ($\beta_{\text{FC}}/beta$), there are similarities in the behavior: $R(\infty)$ is increasing until (in this case) social distancing becomes the prominent strategy. The SIR-SD-FC model with bounded rationality displays oscillations until leveling off below the value for perfect adoption. However, there are also differences: The function now is convex and has a jump discontinuity at the point where face covering becomes ineffective. For any larger value, no measures are taken by the population and the total fraction of infections is equal to the SIR baseline model (if social distancing is too expensive as well). See Figure S1 for an illustration.

\subsection*{Face Covering}

In this part, we want to show that face covering only displays bifurcation behavior if implemented along with social distancing. For this purpose, we look at the SIR-FC model, i.e. the SIR-SD-FC model with $\mathcal{E}(0)=0$. This model displays very similar behavior to the full model. However, when we just change one parameter relating to face covering, there is no upper bound due to social distancing. Interestingly, for any of the parameters ($\beta_{\text{FC}}, C_{\text{I}}, C_{\text{FC}}$), the total fraction of infections $R(\infty)$ displays a jump discontinuity for contagious infections with slow recovery (see Figure S2).

\subsection*{Oscillations in total infections}

Similar to \cite{glaubitz2020oscillatory}, we want to emphasize here again that (i) the SIR-SD-FC model exhibits oscillations in the total infections (unlike the SIR or perfect adoption model) and (ii) the total infections are typically smaller for the model with bounded rationality than for the perfect adoption model. The explanation is related to the concept of herd immunity (HI). HI occurs when enough people have been infected to prevent the spread of infection. After HI is achieved, the total fraction of infected decreases. We refer to the total fraction of infected when herd immunity is reached as $I_{\text{HI}}$.

For $R(\infty)$ to be small, $I_{\text{HI}}$ has to be small. This can be achieved in several ways, like decreasing $I^* = \min(I_3^*,I_1^*)$. So, reducing the perceived cost of social distancing or face covering or increasing the perceived cost of infection or the relative effectiveness of face covering decreases the infections.
To understand the phenomenon of oscillations in $R(\infty)$, we note that reducing $I^*$ in the model with bounded rationality affects $R(\infty)$ in two contradictory ways:
\begin{itemize}
    \item[(i)] Reducing $I^*$ decreases the infections in each wave of infections and thus a decrease in $I_{\text{HI}}$. This causes a reduction in $R(\infty)$.
    \item[(ii)] Once the reduction of $I^*$ passes a threshold, this causes the development of a new wave of infections. Then, $I$ increases again before herd immunity is obtained which in turn increases $I_{\text{HI}}$. Therefore, $R(\infty)$ increases when a new wave of infections develops. 
\end{itemize}

This causes the oscillating behavior that we observe in Figures S1 and S2. When reducing $C_{\text{SD}}$, $C_{\text{FC}}$ or $\beta_{\text{FC}}$ respectively increasing $C_{\text{I}}$, $R(\infty)$ decreases at first (caused by smaller waves of infection and a smaller $I_{\text{HI}}$) followed by an increase (induced by a new wave of infections that causes an increase in $I_{\text{HI}}$).

Additionally, we observe that bounded rationality typically decreases the total infections compared to Perfect Adoption. This again relates to HI. For perfect adoption, herd immunity is obtained for $t = t^*$ with $I = I^*$ while $I_{\text{HI}}$ often is significantly smaller than the model with bounded rationality. If infections are above the threshold when we are close to HI, a small increase in SD or FC can decrease infections below $I^*$. Hence, $I_{\text{HI}} < I^*$ in most cases.

\newpage 
\begin{figure} 
\centering
\includegraphics[width=\textwidth]{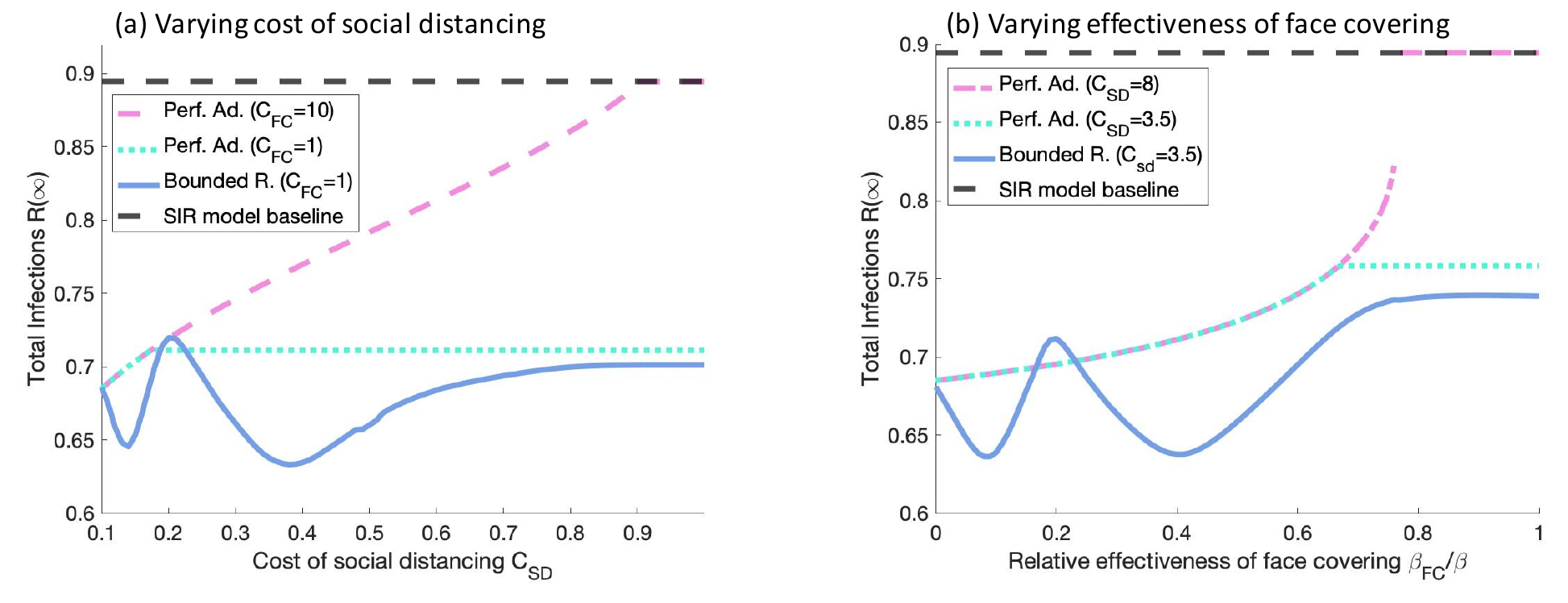} 
\caption{The total fraction of infections $R(\infty)$ for perfect adoption compared to the SIR-SD-FC model. Model parameters are $\beta = 10, \gamma = 4, \omega=3, \kappa=3, C_{\text{I}}=10$.}
\end{figure}

\newpage 
\begin{figure}
\centering
\includegraphics[width=\textwidth]{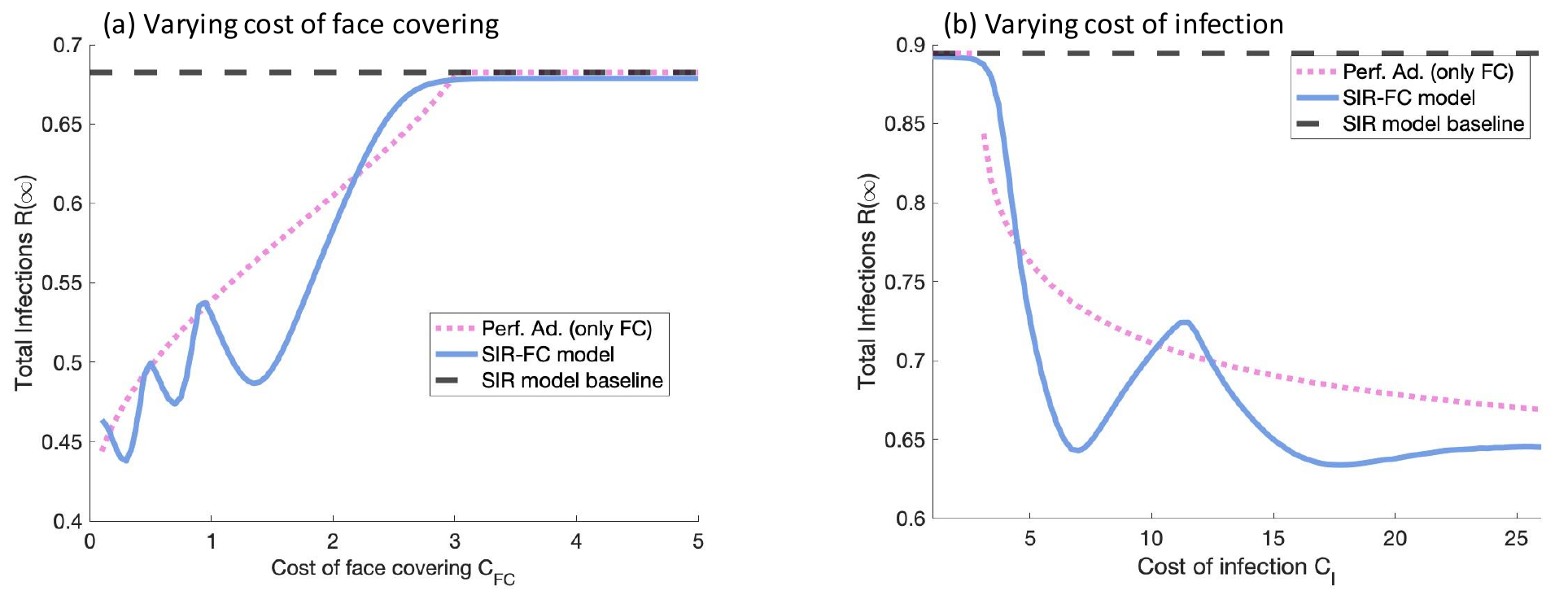}
\caption{The total fraction of infections $R(\infty)$ for perfect adoption compared to the SIR-FC model. Model parameters are (a) $\beta = 10, \gamma = 6, \omega=5, \kappa=5, C_{\text{I}}=10$ respectively (b) $\beta = 10, \gamma = 4, \omega=5, \kappa=5, C_{\text{FC}}=1$.}
\end{figure}




\end{appendices}


\bibliography{sn-bibliography}

\end{document}